\let\jnlstyle=\rmfamily
\def\refjnl#1{{\jnlstyle#1}}%
\newcommand\apj{\refjnl{ApJ}}%
\newcommand\apjl{\refjnl{ApJ}}%
\newcommand\aap{\refjnl{A\&A}}%
\newcommand\icarus{\refjnl{Icarus}}%
\newcommand\mnras{\refjnl{MNRAS}}%

\input{aipcheck}

\documentclass[
    ,final            
  ]
  {aipproc}

\layoutstyle{6x9}

\usepackage{amssymb}
\newcommand{\vct}[1]{\mathbf{#1}}
\newcommand{\cs}{c_\mathrm{s}}
\newcommand{\kmps}{km s${}^{-1}$}
\newcommand{\Msun}{$M_\odot$}

\begin{document}

\title{The Interaction of Planets and Brown Dwarfs with AGB Stellar Winds}


\author{Hyosun Kim}{
  address={Academia Sinica Institute of Astronomy and Astrophysics, 
  P.O. Box 23-141, Taipei 10617, Taiwan; hkim@asiaa.sinica.edu.tw}
}

\author{Ronald E. Taam}{
  address={Department of Physics and Astronomy, Northwestern University,
  2131 Tech Drive, Evanston, IL 60208; taam@tonic.astro.northwestern.edu}
  ,altaddress={Academia Sinica Institute of Astronomy and Astrophysics, 
  P.O. Box 23-141, Taipei 10617, Taiwan; hkim@asiaa.sinica.edu.tw}
}

\begin{abstract}
Beyond the main sequence solar type stars undergo extensive mass loss, 
providing an environment where planet and brown dwarf companions interact
with the surrounding material. To examine the interaction of substellar 
mass objects embedded in the stellar wind of an asymptotic giant branch 
(AGB) star, three dimensional hydrodynamical simulations at high resolution 
have been calculated utilizing the FLASH adaptive mesh refinement code. 
Attention is focused on the perturbation of the substellar mass objects 
on the morphology of the outflowing circumstellar matter. In particular, we 
determine the properties of the resulting spiral density wake as a function 
of the mass, orbital distance, and velocity of the object as well as the wind
velocity and its sound velocity. Our results suggest that future observations
of the spiral pattern may place important constraints on the properties of the 
unseen low mass companion in the outflowing stellar wind.
\end{abstract}

\maketitle


\section{INTRODUCTION}\label{sec:int}

Evolved giant stars lose a significant amount of mass in the form of stellar 
winds, providing material in regions for which orbiting planets or brown 
dwarfs interact with their immediate surroundings. The mass loss rate of 
these giant stars, especially during the asymptotic giant branch (AGB) phase 
of the stellar evolution, ranges from $10^{-7}$ to $10^{-3}$ \Msun~yr${}^{-1}$ 
with a wind speed of 10--30~\kmps\ \citep{fon06}. Such rates lead to 
an envelope density at $\sim$ 100~AU distance from the stellar center of 
$10^{-19}$--$10^{-15}$ $g$~cm${}^{-3}$, approaching the typical density of 
protoplanetary disks \citep{hay81,aik99} where planets form and grow by 
gathering the surrounding matter \citep[e.g.,][]{ali05,hub05,iko00}. The 
gravitational interaction between a protoplanet and its environmental disk 
has been investigated in great detail. Here, a spiral wave forms in the 
differentially rotating system \citep[][and references therein]{gol79,mas08}, 
but the corresponding study in the context of systems interacting with an 
outflowing wind around the evolved giant stars is still lacking. Although 
the velocity direction of the background medium is the only difference between 
the rotating and outflowing systems, the entire background structure and the 
interaction aspects markedly differ. In the former case, where the surrounding 
matter is in the form of a differentially rotating disk, the background 
co-rotates locally with the substellar mass component of interest so that the 
torque at the resonance plays the crucial role in producing the spiral. 
However, in the case of a nearly spherically symmetric stellar wind, the motion
of the substellar object is perpendicular to the background velocity field. 
The relative velocity of the background gas with respect to the substellar 
object causes a Bondi-Hoyle-Lyttleton (BHL) accretion column to lie behind 
the orbiting body \citep[][and references therein]{bon44,edg04}.

Although starting from a different viewpoint, the accretion flow in a weak 
gravitational potential (so-called a density wake) is well formulated by a 
linear perturbation analysis in an initially uniform and static background
\citep{ost99,kim07,kim08}. In particular \citet{kim07} considered an object
moving on a curvilinear orbit, which leads to a density wake in the form of
an Archimedes spiral with the opening angle $\theta=\sin^{-1} (\cs/V_p)$,
where $V_p$ and $\cs$ represent the orbital and sonic speeds. In this 
contribution we show that the wind from the evolved central star pushes the 
wake outward in the radial direction so as to widen the opening of the spiral 
pattern. This provides the possibility for detecting the spiral pattern 
at large distances from the luminous central star and can be used to probe 
the presence of the unseen objects orbiting around the evolved star, if the 
density enhancement is sufficient. To predict the orbital properties of these 
substellar components from future observations of the spiral density wakes, we 
provide an empirical formula for the density enhancement in the spiral pattern.
The reader is referred to \citet{kim11} for a more comprehensive discussion.

\section{SIMULATION SETUP}\label{sec:sim}

Given a {\it continuous} stellar mass loss rate $\dot{M}_\ast$, 
a self-consistent configuration of density $\rho_w$ and velocity $\vct{V_w}$ 
satisfies the steady hydrodynamic conditions
\begin{equation}\label{equ:cow}
  \vct{\nabla}\cdot(\rho_w\vct{V_w})=0,
\end{equation}
and
\begin{equation}\label{equ:mow}
  \vct{V_w}\cdot\vct{\nabla}\vct{V_w}
  = -\frac{1}{\rho_w}\vct{\nabla}P_w-\vct{\nabla}\Phi_\ast
\end{equation}
for the gravitational potential of the central star $\Phi_\ast$
(\citealp{gai87}, \citealp{lam99}, see also \citealp{par58} for the 
original description for isothermal wind solutions). To simplify the problem,
we do not consider radiation pressure on dust coupled with the nearby gas,
which is commonly regarded as the wind driving mechanism of the evolved stars 
\citep[e.g.,][]{win00}. Although the driving mechanism affects the background 
wind condition, the response of gas to the gravitational perturbation due 
to the substellar object orbiting around the mass losing star is likely
independent of the global wind structure as it is determined by the
{\it local} wind quantities. This has been verified by performing two sets of 
simulations with transonic and supersonic branches of \citeauthor{par58}'s 
wind solution, confirming that the wake properties described here are 
determined by the wind density $\rho_w(r)$, its expansion speed $V_w(r)$, 
and the sound speed $\cs(r)$ {\it on the spot} in addition to the orbital 
and accretion properties of the perturbing object. 

We have investigated 42 hydrodynamic models for a range of object mass $M_p$, 
size $r_s$, orbital speed $V_p$, distance from the star $r_p$, as well as 
stellar mass loss rate $\dot{M}_\ast$, wind speed $V_w$, and sound speed 
$\cs$. Specifically, we have verified the dependence of density wake on 
three different velocities: $\cs=3$ and 5~\kmps, $V_p$ is in the range 
of 0.5--10$\cs$, and $V_w$ up to 10$\cs$ satisfying the supersonic and 
transonic branches of isothermal wind models. 
The simulations are performed in three spatial dimensions, 
but here we only focus on the features in the orbital plane. 
See \citet{kim11} for the vertical structures.

\begin{figure}
  \includegraphics[width=\textwidth]{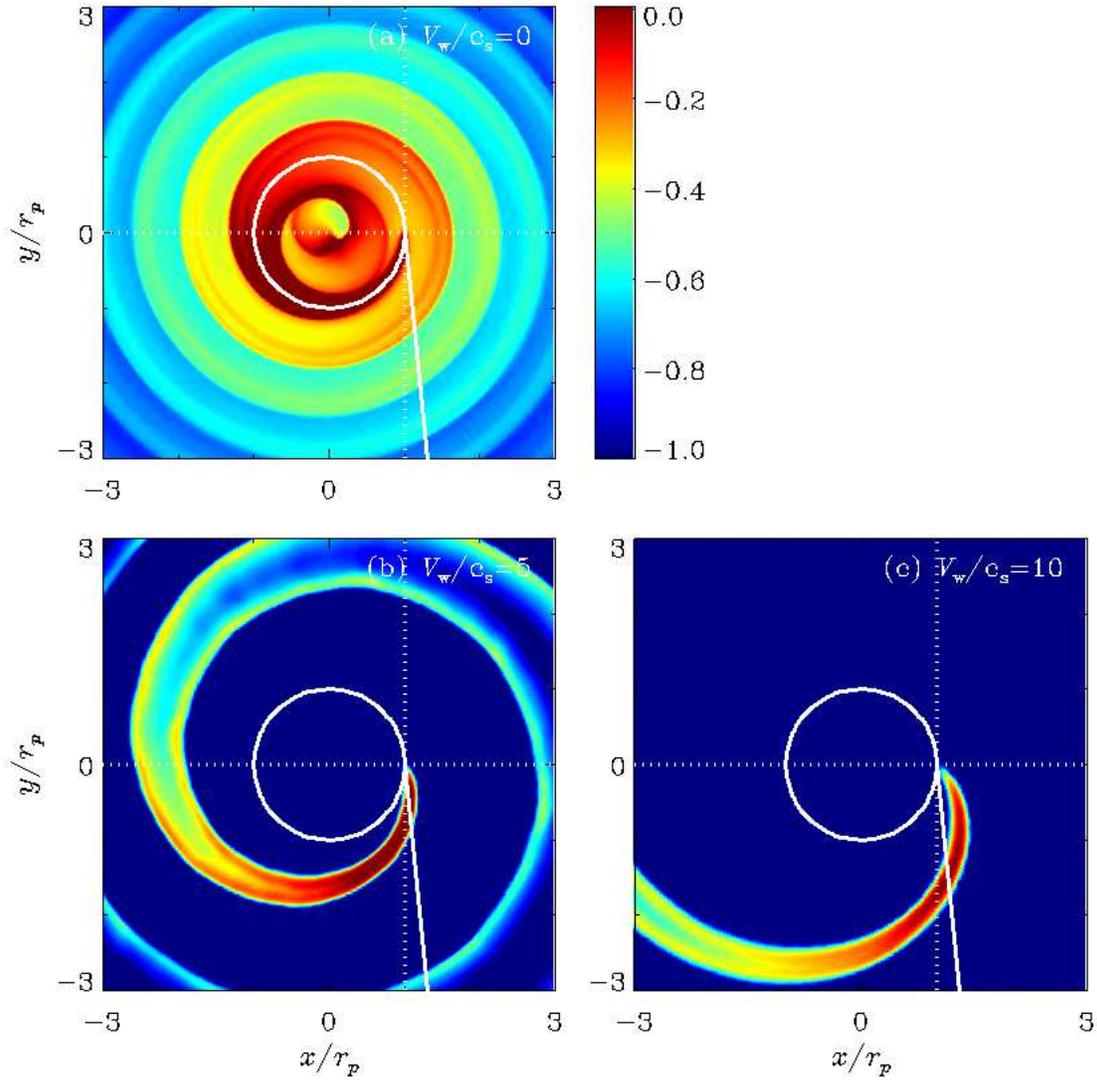}
  \caption{\label{fig:com}
  Comparison of perturbed density distribution between models with the 
  background wind speed $V_w$ of (a) 0, (b) 5, and (c) 10 in the units 
  of sonic speed for a background gas characterized by $\cs=3$~\kmps. 
  The perturbing object is modeled by an object of mass $M_p=0.1$~\Msun\
  moving on a circular orbit at a distance $r_p=20$~AU, marked by 
  a white circle, in the counterclockwise direction with the orbital 
  speed $V_p=10\cs$, currently located at $(x,y)=(r_p,0)$. The white 
  line denotes the opening angle of the spiral in the vicinity of the
  orbiting object in a static medium, $\theta=\sin^{-1} (\cs/V_p)$.
  The color bar labels the perturbed density in logarithmic scale. 
  }
\end{figure}

\section{SHAPE OF SPIRAL PATTERN}\label{sec:pat}

The outflowing wind affects the shape of the spiral density wake, which is 
generated by the gravitational interaction of the substellar object with the 
background. As shown in Figure~\ref{fig:com}, the spiral wake pattern opens
further with higher wind speeds. To show the difference of the opening angle 
of the spiral, one line is overlaid at the present position of the object, 
showing the angle, $\theta=\sin^{-1} (\cs/V_p)$, analytically calculated for 
the model in the absence of a wind \citep{ost99,kim07}. This expression for 
the opening angle reveals that the pattern propagates in the radial direction 
with the sonic speed $\cs$ as the perturbing object moves laterally with 
the speed $V_p$, i.e., the pattern propagation speed is $\cs$ \citep{kim07}. 
Extending this picture to the outflowing models, one can glean from 
Figure~\ref{fig:com} the wind speed is related to the pattern propagation 
speed, which is revealed in the opening angle of the spiral arm pattern. 

Simple geometry yields the following relation
\begin{equation}\label{equ:pat}
\frac{d(r/r_p)}{d\varphi}=\frac{V_{arm}}{V_p},
\end{equation}
from which the pattern propagation speed $V_{arm}$ relative to the orbital 
speed $V_p$ of the perturbing object can be derived from the shape of the 
spiral. Numerical differentiation at the shock boundary, exhibiting high
density, gives the pattern propagation speed of $V_w+\cs$, which is also
predicted by linear analysis in the WKB approximation. Here, we point out 
that it is not well described purely by the speed of outflowing matter, 
which is often used by observers \citep[e.g.][]{mau06,mor06}, especially 
when the wind speed is comparable to the thermal sound speed. 
Also it is to be noticed that the pattern is not necessarily an Archimedes 
spiral, in which the arm spacing is constant, but its shape possibly changes 
with the variation of wind speed and temperature as a function of distance. 

\section{DENSITY ENHANCEMENT}\label{sec:jum}

To detect the spiral arm pattern around the evolved giant stars, the density 
enhancement (or density jump) of the structure must exceed some threshold 
value. In our definition $\alpha=\rho/\rho_w-1$ represents this perturbed 
density normalized by the {\it local} background density $\rho_w (r)$, 
which has the profile close to $\sim r^{-2}$ in most areas.

\begin{figure}
  \includegraphics[width=\textwidth]{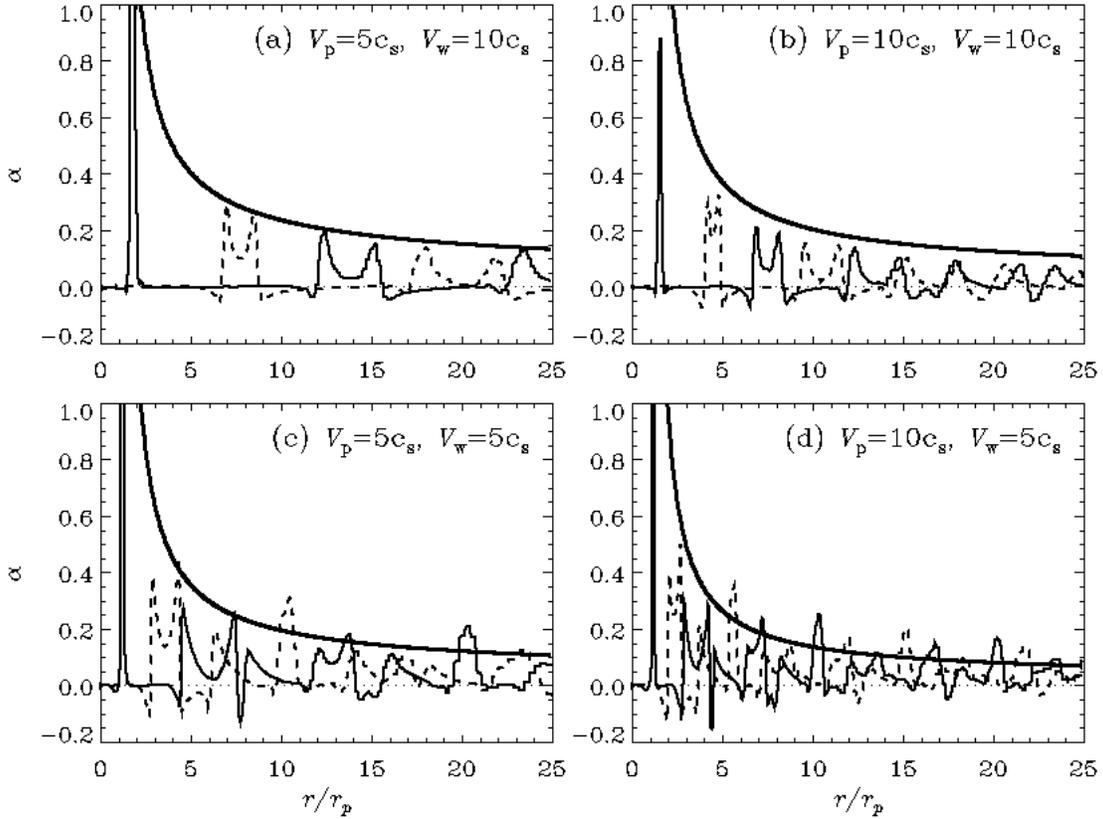}
  \caption{\label{fig:jum}
  Density enhancement $\alpha$ of the spiral density wake formed by the 
  gravitational interaction of the stellar wind with a substellar object 
  of mass 0.1~\Msun\ orbiting around the central star as a function 
  of distance normalized by the orbital radius $r_p=20$~AU. 
  Comparison is achieved between the models with the orbital and wind 
  speeds of (a) $V_p=5\cs$, $V_w=10\cs$, (b) $V_p=10\cs$, $V_w=10\cs$, 
  (c) $V_p=5\cs$, $V_w=5\cs$, and (d) $V_p=10\cs$, $V_w=5\cs$, where 
  the sound speed $\cs$ is 3~\kmps. The lines denote the perturbed density 
  profiles along the distance from the star in the direction of the object 
  ({\it solid}) and in the opposite direction ({\it dashed}).
  See text for the complementary bold solid line that outlines the peaks.
  }
\end{figure}

Figure~\ref{fig:jum} exhibits the density enhancement for four different models
with different combinations of the orbital speed and the wind speed but with
a fixed sound speed, as well as the mass, radius, and orbital radius of the 
object, and the stellar mass and mass loss rate. We have also checked the 
effects of all other parameters enumerated here to obtain an empirical formula 
for the minimum value of the density peaks at the shock boundary of the spiral 
arm:
\begin{equation}\label{equ:jum}
\alpha_{peak}\gtrsim\frac{r_B}{|r-r_p|}\left(\frac{
\frac{|r-r_p|}{r_B}V_w^2 + 10V_p V_w + \cs^2}{V_w^2+|V_p^2-\cs^2|}\right)^{1/2}
\qquad {\rm for}\ \ V_w>\cs,
\end{equation}
where $r_B=GM_p/\cs^2$ is the Bondi accretion radius. We mention that 
this empirical formula can be improved by exploring additional parameter 
space and/or by direct analytic approach, although it serves quite well 
for a large range of parameter space as the lower limit of density peaks 
that could be detected. Note that this equation is well checked only for the 
cases of a supersonic speed of wind, even though the density wakes in the 
transonic wind background appear reasonable in the inner subsonic regime. 

For a fast wind, which is often observed in AGB envelopes, 
this equation~(\ref{equ:jum}) is reduced simply to 
\begin{equation}\label{equ:jfw}
\alpha_{peak}\gtrsim\left(\frac{r_B}{r}\right)^{1/2}
\qquad {\rm for}\ \  V_w\gg V_p,\,\cs,
\end{equation}
when the observed arm is located at great distances from the central star
relative to the orbital radius of the perturbing object. In comparison to 
the slow wind cases, the density peak profiles decrease with distance 
from the star with a power of $-0.5$ rather than $-1$. This is favorable
for observers seeking a signature of unseen low mass objects revolving
about evolved mass losing giant stars.

\section{DISCUSSION}\label{sec:dis}
\subsection{Prediction of the Orbital Properties of Planets and Brown Dwarfs 
Deduced from Spiral Structures of AGB Stellar Wind Envelopes}

Equation~(\ref{equ:jfw}) indicates that for the typical envelope temperature 
of the cool giants, corresponding to the sound speed of 1--4~\kmps, 
the peak density jump for objects of 1~$M_J$, 30~$M_J$, and 0.1~\Msun\ is 
greater than 2.5--10\%, 14--55\%, and 25--100\%, respectively, at 100~AU
distance from the star. Although such enhancements are not large, they may 
be detectable in the near future with sufficiently sensitive observations. 
In the fast wind assumption, the mass of the unseen object may be estimated
by measuring the density jump of the spiral arm at a given position, 
given knowledge of the sound speed. Furthermore, if more than one turn of 
the spiral is detected, the orbital properties could be also estimated
by fitting the density peak profile.

In this case that more than one turn are detected, 
additional information can be inferred from the arm interval. Assuming 
the pattern propagation speed does not vary significantly along one turn,
the integration of equation~(\ref{equ:pat}) gives the arm interval, defined 
by the distance between the outer boundary of the pattern, as following:
\begin{equation}\label{equ:dr1}
\Delta r_{arm}=(V_w+\cs)\times\frac{2\pi r_p}{V_p}.
\end{equation}
From this result, one can estimate the orbital period of the companion 
provided that $\Delta r_{arm}$, $V_w(r_{arm})$, and $\cs(r_{arm})$ are known
from multi-wavelength observations. Alternatively, it can be rewritten as
\begin{equation}\label{equ:dr2}
\Delta r_{arm}=\frac{2\pi(V_w+\cs)}{(GM_\ast)^{1/2}}\times r_p^{3/2},
\end{equation}
which can be used to determine the orbital distance of the companion 
provided that knowledge of the stellar mass $M_\ast$ is obtained separately.

The application of our results to the analysis of observation data is somewhat
qualitative given that there are many uncertainties in the inner part of the 
envelopes resulting from the effects associated with the detailed driving of
the stellar wind by stellar radiation, pulsation of the AGB star, the gas 
heating-cooling properties, and the accretion properties of the substellar 
object. Nevertheless our results suggest that the search for spiral arm 
patterns in the circumstellar envelope of an AGB star is desirable in order to
make use of the full potential of the probes described in this paper to infer
the presence and to constrain the properties of unseen objects orbiting about 
evolved mass-losing stars.


\begin{theacknowledgments}
This research is supported by the Theoretical Institute for Advanced 
Research in Astrophysics (TIARA) in the Academia Sinica Institute of 
Astronomy and Astrophysics (ASIAA). 
We would like to thank the organizing committee of the workshop on Planetary 
Systems Beyond the Main Sequence for encouraging the presentation as well as 
the workshop participants for valuable discussion.
H.K. is grateful to Paul M. Ricker for advice on computational issues.

\end{theacknowledgments}

\bibliographystyle{aipproc}   




\end{document}